\documentclass[a4paper,fleqn,usenatbib]{mnras}

\usepackage[T1]{fontenc}
\usepackage{ae,aecompl}
\usepackage{graphicx}
\usepackage{bm}
\usepackage{amssymb}
\usepackage{amsmath}
\usepackage{amsfonts}
\usepackage{dcolumn}
\usepackage{natbib}
\usepackage{color}
\usepackage{calc}
\usepackage[normalem]{ulem}
\usepackage{verbatim}

\def\apj{Astrophys. J.}
\def\apjl{Astrophys. J. Lett.}
\def\apjs{Astrophys. J. Supp. Ser. }

\def\aap{Astron. Astrophys. }

\def\mnras{Mon. Not. Roy. Astron. Soc. }

\def\prd{Phys. Rev. D.}

\def\Dwa{$\,$\uppercase\expandafter{\romannumeral5}$\,$}

\def\sless{\lower2pt\hbox{$\buildrel {\scriptstyle <}
   \over {\scriptstyle\sim}$}}

\def\sgreat{\lower2pt\hbox{$\buildrel {\scriptstyle >}
   \over {\scriptstyle\sim}$}}
\def\sharpnull#1{}

\newcommand{\code}[1]{\texttt{#1}}

\let\a=\alpha \let\b=\beta  \let\d=\delta 
  \let\th=\theta  
 \let\m=\mu \let\n=\nu  
 \let\t=\tau  \let\f=\phi

 \let\P=\Pi \let\S=\Sigma  
 
\let\eps=\varepsilon

\let\ci=\citep
\def\no{\nonumber} 
 \def \size {\displaystyle}
\newcommand{\rf}[1]{(\ref{#1})}

\def \del {\partial}

\def\ba{\begin{array}} \def\ea{\end{array}}
\def \be {\begin{eqnarray}} \def\ee{\end{eqnarray}}
\def \bei {\begin{itemize}} \def \eei {\end{itemize}}

\def \ben {\begin{equation*}}
\def \een {\end{equation*}}

\def \f {{\bf f}}
\def \P {\hat{P}}

\def \farc {\frac}

\def \f {{\cal F}}

\title{Analytic Closures for M1 Neutrino Transport}

\author[E. M. Murchikova et al.]{
E. M. Murchikova$^{1}$\thanks{lena@tapir.caltech.edu},
E. Abdikamalov$^{2}$,
T. Urbatsch$^{3}$
\\
$^{1}$TAPIR, MC 350-17, California Institute of Technology, 1200 E California Blvd., Pasadena, CA 91125, USA\\
$^{2}$Department of Physics, School of Science and Technology, Nazarbayev University, Astana 010000, Kazakhstan\\
$^{3}$Los Alamos National Laboratory, Los Alamos, NM, USA
}

\begin{document}
\label{firstpage}
\pagerange{\pageref{firstpage}--\pageref{lastpage}}
\maketitle

\begin{abstract}
Carefully accounting for neutrino transport is an essential component of many astrophysical studies. Solving the full transport equation is too expensive for most realistic applications, especially those involving multiple spatial dimensions. For such cases, resorting to approximations is often the only viable option for obtaining solutions. One such approximation, which recently became popular, is the M1 method. It utilizes the system of the lowest two moments of the transport equation and closes the system with an ad hoc closure relation. The accuracy of the M1 solution depends on the quality of the closure. Several closures have been proposed in the literature and have been used in various studies. We carry out an extensive study of these closures by comparing the results of M1 calculations with precise Monte Carlo calculations of the radiation field around spherically-symmetric protoneutron star models. We find that no closure performs consistently better or worse than others in all cases. The level of accuracy a given closure yields depends on the matter configuration, neutrino type, and neutrino energy. Given this limitation, the maximum entropy closure by Minerbo (1978) on average yields relatively accurate results in the broadest set of cases considered in this work. 
\end{abstract}


\section{Introduction}

Neutrinos play an important role in core-collapse supernovae (CCSNe), coalescence of binary neutron stars and many other astrophysical phenomena. Their collective behavior is described by the distribution function that obeys the Boltzmann transport equation. The state of the radiation field is characterized by spatial coordinates, the propagation direction (two angles), energy, and time, making the problem seven-dimensional in the most general case. Many astrophysical systems have dense and opaque central regions surrounded by transparent low-density envelopes. Radiation moves within the dense central regions via diffusion and, when it leaks into the outer regions, it streams freely. The transport equation has a parabolic character in the former region, while in the latter region, it has a hyperbolic character \cite[e.g.,][]{mihalas:84,pomraning:83}. In order to model such systems accurately, the solution techniques must handle not only the two different regimes, but also the transition between the two. In the presence of scattering, the collision terms on the right-hand side of the Boltzmann equation contains the angular moments of the specific intensity, which makes the Boltzmann equation an integro-differential equation. These aspects make solving the transport equation a challenging computational task. For this reason, often one has to resort to approximations and simplifications to make the problem tractable. 

One way of simplifying the problem is to assume spherical or axial symmetries to reduce the number of spatial dimensions. There are many situations where such assumptions have been employed. Spherically-symmetric calculations have been performed by, e.g., \citet{mezzacappa:93a,mezzacappa:93b,mezzacappa:93c,yamada:97,ramppjanka:00,liebendoerfer:05,sumiyoshi:05}. 2D axisymmetric simulations have been performed by, e.g., \cite{livne:04,ott:2008,brandt:11,skinner:16,burrows:16}. In addition, ray-by-ray approaches, in which multi-dimensional transport problem is approximated as a set of one-dimensional problems along radial rays, has been used widely \citep[e.g.,][]{bhf:95,buras:06a,marek:09,mueller:11a,mueller:12a,bruenn:13,bruenn:16,lentz:15,melson:15a,melson:15b,mueller:15,mueller:15b}.
Most realistic problems, however, do not possess spatial symmetries. For these problems, the transport equation has to be solved in full three dimensions. The pioneering attempts to solve three-dimensional Boltzmann equation have been already taken \cite[e.g.,][]{radice:13rad,sumiyoshi:12,sumiyoshi:15}, yet the computational cost remains too high for solving it in realistic scenarios. 

In order to further reduce the cost, one can approximate the Boltzmann equation either by neglecting the time dependence ({\it steady-state solution}) and/or energy dependence ({\it gray schemes}). The simplest treatment of the transport problem is the "light bulb" approach, in which simple parametrized neutrino heating and cooling rates are imposed \citep[e.g.,][]{ohnishi:06,murphy:08,radice:16a}. The less cruder approximation, the so-called leakage/heating scheme, has been used extensively in the literature \citep[e.g.,][]{ruffert:96,rosswog:03b,oconnor:11,ott:12a,ott:13neutrino,ott:13a,moesta:14b,abdikamalov:15,deaton:13}. The leakage/heating scheme evaluates the local neutrino energy and number emission rates, which are then locally subtracted from the matter. A fraction of the emitted energy is deposited back as neutrino heating in the ``gain'' region outside the protoneutron star (PNS)\citep[e.g.,][]{oconnor:10,ott:13a}.

In this paper we focus on an alternative approach, employing the reduction of the angular degrees of freedom of the problem, called the {\it moment scheme}. The simplest version of the moment scheme is the diffusion approximation. One takes the zeroth moment of the transport equation, which yields an equation that contains the zeroth and first moments of the distribution function. For non-static moving media, the third moment is also present \cite[e.g.,][]{just:15b}. These three moments represent the energy density, flux, and pressure tensor of radiation, respectively. In the optically thick limit, the first moment can be approximated using the gradient of the zeroth moment via the Fick's law, while the second moment can be approximated as one-third of the zeroth moment \cite[e.g.,][]{pomraning:83}. These relations allows us to ``close'' the zeroth moment of the transport equation. While the resulting equation is far simpler than the original Boltzmann equation, the diffusion approximation is not valid in the free-streaming regime and yields inaccurate results such as acausal flux. This can be mitigated by using the flux limiter \citep[e.g.,][]{smit:00,burrows:00} or by using other advanced prescriptions \citep[e.g.,][]{janka:92b,scheck:06,mueller:15}. The way to obtain a more accurate solution is to incorporate higher-order moments. 

The first moment of the transport equation results in an equation containing up to the second moments of the distribution function. In general relativity and for non-static media, the third moment is also present \citep{shibata:11,cardall:13rad,just:15b}. Together with the zeroth moment of the transport equation the system has four sets of unknowns: the zeroth, first, second, and third moments. There are two commonly used approaches for closing the system. In the first method, using the first and second moments as given, one can express the source terms of the Boltzmann equation due to interaction with matter as functions of only space, time, and momentum coordinates \citep{burrows:00,rampp:02}. This transforms the Boltzmann equation from a non-linear integro-differential equation into a linear differential equation. The solution of this simpler equations yields higher moments to close the original system of the lowest two moment equations, from which we can obtain updated zeroth and first moments. This procedure is iterated until convergence. Depending on the method for obtaining the closure, the approach can yield the full solution of the Boltzmann equation. This method is usually called the variable Eddington tensor method \citep{burrows:00,rampp:02,cardall:13rad}. 

Another approach for closing the system is by expressing the second and the third moments in terms of the lower-order moments using {\it approximate} analytical relations. This results in a closed system of two equations for the zeroth and first moments. Originally proposed by \citet{pomraning:69, kershaw:76,levermore:84}, such methods are often called the M1 methods \citep{smit:97,pons:00,audit:02} or the ``algebraic Eddington factor'' methods \citep{just:15b}. 

A common way to derive a closure relation by interpolating between optically thick and optically thin limits. In these limits, the second and third moments can be expressed precisely in terms of the zeroth and first moments\footnote{More specifically, in the optically thin limit, one can derive an expression for the second and third moments for a freely propagating radiation beam \cite[e.g.,][]{shibata:11}.}. The Eddington factor, which ranges from $1/3$ to $1$ between these limits, serves as an interpolation parameter. The functional form of the Eddington factor in terms of the local energy density and flux of radiation is a relation often called as the M1 closure in the literature. Once this relation is established, the system is closed \cite[e.g.,][]{shibata:11}\footnote{This is not the only approach for closing the system that exists in the literature. In principle, if the second and third moments are assumed to be functions of the local values of the lowest two moments, then the second moment can be expressed in terms of the latter two via the Eddington factor \citep[e.g.,][]{just:15b}. Similarly to the second moment, the third moment can be expressed in terms of the lowest two via the "third-order counterpart" of the Eddington factor. To close the system, this factor must be expressed in terms of the radiation energy density and flux, which has been achieved by, e.g., \citet{vaytet:11} and \citet{just:15b} for two different closures.}.

The M1 approach is particularly suitable for problems with not too complex geometries such as CCSNe and remnants neutron star mergers. In these problems, the radiation field is often arranged in such a way that there exists some approximate relationship between higher-order and lower-order moments. That said, not all problems possess such properties. A prominent example is a collision of radiation beams coming from multiple sources. For this problem, the closure relation depends not only on the local values of the first and second moments, but also on the spatial distribution of the radiation sources. In general, if the second and third moments are assumed to be functions of the zeroth and first moments, then the former two must be symmetric with respect to rotation around the direction of the radiation flux \citep[e.g.,][]{cardall:13rad}. For problems with such symmetries, the M1 approach offers excellent compromise between computational cost and accuracy.

Moreover, the moment equations constitute a hyperbolic system, which allows us to utilize a wide variety numerical methods developed for solving hyperbolic system of conservations laws (e.g., Godunov-type methods) to calculate the solution of the transport problem \citep{pons:00}. For these reasons and because of their relatively modest computational cost, such methods recently gained significant popularity in astrophysics.

The M1 method has been applied to a wide range of problems such as core-collapse supernovae \cite[e.g.,][]{oconnor:15b,roberts:16c,kuroda:16}, evolution of protoneutron stars \citep{pons:00}, accretion disks \cite[e.g.,][]{shibata:12,foucart:15,just:15b,just:16}, and many more in AGN accretion literature. A number of analytical closures have been suggested in the literature. The accuracy of the M1 solution depends on the quality of the closure used and it is a priori unclear which closure yields the best results for a given problem. While the quality of individual closures has been examined in different contexts \cite[e.g.,][]{janka:91phd,smit:97,oconnor:15a}, a systematic analysis for neutrino transport has been performed only by \citet{janka:91phd}, \citet{smit:00}, and \citet{just:15b}\footnote{Note that there is a relation between the M1 scheme and the flux-limited diffusion approximation and each flux limiter is associated with an M1 closure relation \citep{levermore:84,smit:00}. The quality of some of the flux limiters has been explored by, e.g., \citet{burrows:00,just:15b} using the flux-limited diffusion framework for neutrino transport in the context of core-collapse supernovae.}. The aim of this work is to extend these two works, to consider a wider selection of M1 closures, to verify them using a wider range of test problems that are relevant to neutrino transport, and to present a quantitative assessment of their quality.

In this work, we evaluate the quality of various closures proposed in the literature by comparing the radiation field distribution in and around radiating objects obtained with the M1 method with the one obtained analytically or with Monte Carlo. We consider two types of radiating objects: a uniform sphere with a sharp surface and a protoneutron star with a hot envelope obtained from core-collapse simulations. These two objects possess the opaque central radiating source surrounded by a transparent envelope, i.e. the important characteristics common to many astrophysical systems. Since our goal is to study the quality of the analytical closures and in order not to contaminate our results with errors emanating from other sources such as hydrodynamics and non-linear radiation-matter coupling, we consider only static matter configurations in our tests. For simplicity, we limit ourselves to spherical symmetry and ignore spacetime curvature around PNSs. Implications of these assumptions will be discussed in Section~\ref{sec:conclusion}.

The uniform sphere problem consists of an opaque radiating sphere with a sharp surface surrounded by vacuum, and it has an analytical solution (cf. Section~\ref{sec:sphere}). In the PNS problem, we take three different post-bounce configurations (obtained from simulations of \citealt{ott:2008}) of a $20M_\odot$ progenitor star at 160, 260, and 360 ms after bounce. We obtain precise solution of this problem by performing Monte Carlo radiation transport calculations using the code of \cite{abdikamalov:12} (cf. Section~\ref{sec:pns}). These solutions are compared to M1 solutions obtained using the GR1D code \ci{oconnor:11,oconnor:13,oconnor:15a} available at \url{http://www.GR1Dcode.org}. 

We consider seven different closures. These are the closures by \citet{kershaw:76}, \citet{wilson:75}, \cite{levermore:84}, the classical maximum entropy closure of \citet{minerbo:78}, and the maximum entropy closure with the Fermi-Dirac distribution by \citet{cernohorsky:94}. In addition, we consider two closures by \citet{janka:91phd} that are constructed by fitting closure relations to exact Monte Carlo solutions of the radiation field around PNSs \citep{janka:91phd}. 

This paper is organized as follows. In Section \ref{sec:theory}, we give a theoretical overview of the neutrino transport problem and the M1 scheme. In Section \ref{sec:closures}, we give a brief description of the seven closures we study in this work. Section \ref{sec:results} presents the details of the test problems. We also describe our tools for systematic quantitative assessment of the quality of the closure relations and present the results of our analysis. Our conclusions are provided in Section \ref{sec:conclusion}. In Appendix \ref{sec:methods}, we describe the two codes that we use in our analysis: the {\tt GR1D} code for M1 transport and the Monte Carlo code of \citet{abdikamalov:12}.

\section{Boltzmann Equation and M1 method}
\label{sec:theory}

Neutrinos are described by the distribution function $\f$, which characterizes the number of neutrinos in a phase-space volume element and which obeys the relativistic Boltzmann equation \citep[e.g.,][]{lindquist:66,mezzacappa:89}:
\be
        \frac{d x^\a}{d \t} \frac{\del \f}{\del x^\a}+
        \frac{d p^i}{d \t} \frac{\del \f}{\del p^i}=
        (-p^\a u_\a) S(p^\m, x^\m, \f)\,\,.
\ee
Here, $\t$ is an affine parameter of the neutrino trajectory, $u^\m$ is the four-velocity of the medium, and  $p^\m$ is the four-momentum of radiation, from which one can obtain the neutrino energy in the rest frame of the medium via relation $\varepsilon = -p^\a u_\a$. The Greek indices $\m, \a = 0, 1, 2, 3$ run over space-time components and the Latin indices $i = 1, 2, 3$ runs over the spatial components. $S(p^\m, x^\m, \f)$ is the collision term that describes the interaction of radiation with matter via absorption, emission and scattering. The evaluation of $S(p^\m, x^\m, f)$ is a domain of a separate field of study and it is beyond the scope of this work \citep[e.g.,][]{bruenn:85,brt:06}. In this work, we treat neutrinos as massless particles and fix units using $\hbar = c = 1$. 

The zeroth, first, and second moments of the distribution function represent the energy density,
\be\label{M0e}
    E_\nu=\int \eps \f(p^\m,x^\m) \delta(h\nu-\eps) d^3p\,,
\ee
the radiation flux,
\be\label{M1f}
    F^j_\nu=\int p^j \f(p^\m,x^\m) \delta(h\nu-\eps) d^3p\,,
\ee
and the radiation pressure,
\be\label{M2p}
    P^{ij}_\nu=\int p^i p^j \f(p^\m,x^\m) \delta(h\nu-\eps)
        \frac{d^3p}{\eps}\, .
\ee
Here, $E_\nu$, $F^j_\nu$, and $P^{ij}_\nu$ are the functions of neutrino energy $\eps=p^0=|\vec{p}|$. In order to obtain the total energy density, flux, and pressure, one has to integrate (\ref{M0e})-(\ref{M2p}) over energy, as discussed in, e.g., \cite{thorne:81, novikov:73}.

The zeroth and the first moments of the Boltzmann equation constitute the system of equations for $E_\nu$ and $F^j_\nu.$ In Minkowski space, spherical symmetry, and neglecting the velocity of the medium, these two evolution equations are
\be\label{eq1}
    && \frac{\del}{\del t} E_\n + \frac{1}{r^2} \frac{\del}{\del r} \left( r^2 F_\n^r \right) 
= S[0]_{\nu}\\
    && \frac{\del}{\del t} F_\n^r + \frac{1}{r^2} \frac{\del}{\del r} \left( r^2  P^{rr}_{\n}\right) 
= S[1]^{r}_{\nu}\label{shm},\label{eq2}
\ee
where $S[0]_{\nu}$ and $S[1]_{\nu}$ are accordingly the zeroth and first moments of the collision term  $S(p^\a, x^\b, \f)$. Note that, since we consider flat spacetime and static matter, the third moment does not appear in this equation. As we pointed out above, this system is not closed. There are two equations (\ref{eq1}) and (\ref{eq2}), but three unknowns $E_\nu$, $F^r_\nu$, and $P^{rr}_\nu$. This is a simple reflection of the fact that, although the system (\ref{eq1})-(\ref{eq2}) is obtained from the Boltzmann equation, it does not capture all the information contained in the Boltzmann equation. To capture the complete information, one has to solve the complete system, which can be expressed as 
\be\label{eqk}
    {\rm Function}(M[0] ... M[k+1])=S[k],
\ee
where
\be\label{Mk}
        M[k]^{\a_1 ... \a_k}_{\eps_0}=\int \eps^2 \f(p^\m,x^\m) \delta(h \nu-\eps)
        \frac{p^{\a_1}}{\eps}... \frac{p^{\a_k}}{\eps} \frac{d^3 p}{\eps},
\ee
is the moment of order $k$. Note that subscripts and superscrips are omitted in this equation to avoid clutter. This is an infinite system of an infinite number of unknowns $M[k]$, which is not feasible to solve in practice. 

This situation is somewhat analogous to the Taylor expansion. Function $f(x)$ can be represented through
the infinite sum
\be
f(x)= \sum\limits_N\frac{1}{N!}\frac{d^N f(x)}{dx^N}\bigg|_{x=x_0}(x-x_0)^N.
\ee
This allows one to express the value of $f(x)$ at an arbitrary point $x$ through its properties at a given point $x_0$. In order to calculate $f(x)$ exactly, one has to incorporate all the terms in the infinite series. The low-order terms yield accurate results only in the vicinity of the point $x_0$. Similar is true 
for the moments of the distribution function. When we constrain ourselves to the first few moments, we sacrifice the accuracy of our description. To capture all the information contained in the distribution function one  needs to employ the whole infinite set of moments.

The M1 approach used in the literature is based on an interpolation of the radiation pressure $P^{ij}$ between optically thick and thin limits \citep[e.g.,][]{shibata:11}
\be\label{eq:closure} 
	&& P^{ij}_\n= \frac{3 p -1}{2} P^{ij}_\mathrm{thin}+\frac{3(1-p)}{2} P^{ij}_\mathrm{thick},
\ee
where $P^{ij}_\mathrm{thick}$ and $P^{ij}_\mathrm{thin}$ are the radiation pressure in these limits. In the former limit, radiation is in thermal equilibrium with matter and is isotropic. This results in $F_\nu^i=0$ and 
\be
	P^{ij}_\mathrm{thick}=\frac{1}{3}E_\nu \d^{ij}
\ee	
for the gas of ultrarelativistic particles such as photons and neutrinos \citep{mihalas:99}. In the free-streaming limit, radiation propagates like a beam along a certain direction $n$ and exerts pressure only along this direction, giving us $F^{n}_\nu = E_\nu$ and $F^{i\neq n}_\nu = 0$ and
\be
	P_\mathrm{thin}^{nn} = E_\n \frac{F_\n^n F_\n^n}{|F_\nu|^2}, \quad {\rm and} \quad P_\mathrm{thin}^{ij}=0, \, 
    {\rm if} \, i \, {\rm or} \, j \neq n.
\ee	
The parameter $p$ in equation (\ref{eq:closure}) is known as the variable Eddington factor and it plays the role of the interpolation factor between the two regimes. The functional form of $p$ in terms of the lower moments is referred to as the M1 closure.

Equation (\ref{eq:closure}) is derived based on the assumption that the radiation is symmetric around the direction parallel to the flux. While the assumption is valid for the spherically symmetric matter and radiation distributions, it does not always hold in more general cases. Colliding radiation beams emanating from different sources is a prominent example. Therefore, equation (\ref{eq:closure}), even before we fix the form of the function $p$, already contains an approximation. 

Note that equation (\ref{eq:closure}) in its modern form is often cited as derived by \cite{levermore:84} in the literature. While it is true, \cite{kershaw:76} also proposed the interpolation between thick and thin regimes like (\ref{eq:closure}). He then suggested using the simplest among such interpolation
\be\label{pr}
	P_\nu^{ij} = E f^i f^j + \frac{E}{3} \d^{ij} \left(1-f^2 \right),
\ee
where 
\be
f^i = F_\nu^i/E_\nu
\ee
and $f^2 = f^i f_i$. This relation is equivalent to equation (\ref{eq:closure}) with a closure relation $p = (1+2 f^2)/3$, which is known as the Kershaw closure. Even earlier, a formulation similar to M1 was discussed by \citet{pomraning:69}.

\section{Closures}
\label{sec:closures}

In this section, we present a list of seven different closures most commonly used in the literature and describe their main properties. 

\begin{figure*}
\centering
\includegraphics[width=16cm]{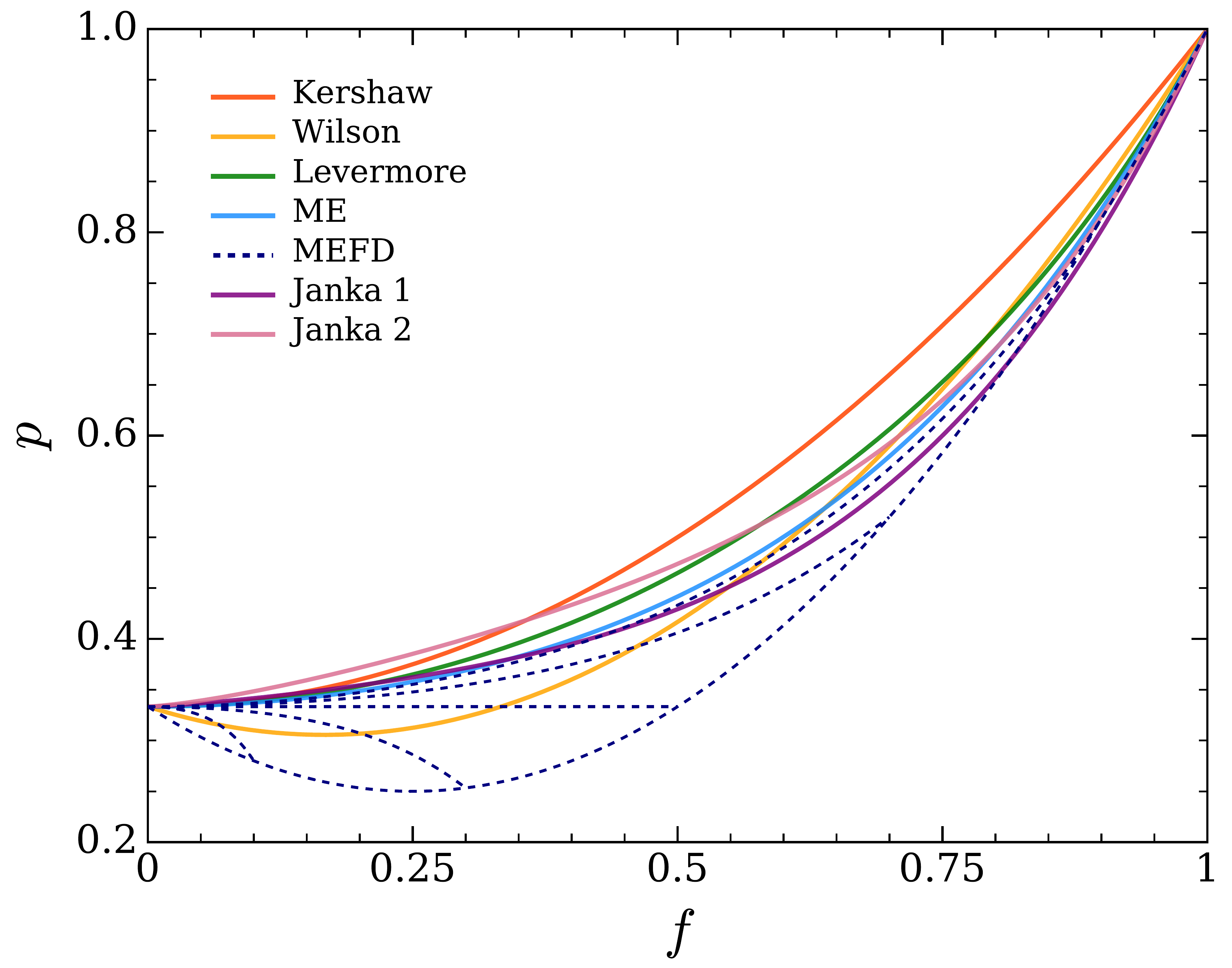}
\caption{The closure relations for the Eddington factor $p = P^{rr}_\n/E_\n$ as the function
of flux factor $f = F^r_\n /E_\n.$ The {\tt MEFD} closure is a two parameter function and is 
represented by series of curves for $e=0.1,\ 0.2,\ 0.3,\ 0.5,\ 0.7$, and $0.9$. The bottom curve is the limit of maximal packing. In the limit $e\rightarrow 0$, the {\tt MEFD} closure reduces to its classical limit, the {\tt ME} closure, represented by the solid sky blue line.  
(see Section \ref{MEFD}).\label{fig:closures}} 
\end{figure*}

\subsection{Kershaw Closure}
The \citet{kershaw:76} closure is a simple interpolation between the optically thick ($f \to 0$) and the optically thin ($f \to 1$) limits. In the spherically symmetric, case the Kershaw closure reads
\be\label{kshw}
	p = \frac{1+2 f^2}{3}.
\ee
This closure is shown with the solid red line in Fig.~\ref{fig:closures}. In the following, we refer to this closure as the {\tt Kershaw} closure. 

\subsection{Wilson Closure}

\citet{wilson:75} and \citet{leblanc:70} presented a flux-limiter for neutrino diffusion, which corresponds to the closure 
\be
	p=\frac{1}{3}-\frac{1}{3} f + f^2.
\ee
Physically, this expression is equivalent to an interpolation of the diffusive and free-streaming fluxes via harmonic averaging \citep{smit:00}. This ensures correct diffusive and free-streaming limits, but may yield imprecise results in the intermediate regime. This closure is shown with the solid yellow line in Fig.~\ref{fig:closures}. Hereafter, we refer to this closure as the {\tt Wilson} closure. 

\subsection{Levermore Closure}

The Levermore closure can be derived assuming that the radiation is isotropic and satisfies the Eddington closure ($P^{ij}_{\nu}=P^{ij}_{\rm thick}$ or $p=1/3$ everywhere) in the ``rest frame'' of radiation, i.e., in the frame in which the radiation flux is zero \citep{levermore:84, sadowski:13}:
\be
	p=\frac{3+4 f^2}{5+2\sqrt{4-3 f^2}}.
\ee
This closure is shown with the solid green line in Fig.~\ref{fig:closures}. We refer to this closure as the {\tt Levermore} closure.

\subsection{MEFD: Maximum Entropy Closure for Fermionic Radiation}\label{MEFD}

The idea to use the maximum entropy principle to construct the closure relation was first suggested by \citet{minerbo:78}, who applied it to photons assuming a classical distribution. Later, \citet{cernohorsky:94} applied it to fermions using the Fermi-Dirac distribution. 

By maximizing the entropy functional 
\be
S[\f(\mu)]= (1-\f)\log(1-\f)+\f \log \f,
\ee
under the constraints that the dimensionless zeroth and first moments,
\be\label{ef}
	e&=&\frac{E_\nu}{\n^3}=\int\limits^{2\pi}_{0}d\phi  \int\limits^{1}_{-1} \f d \mu,\\
	f&=&\farc{F_\nu}{E_\nu}=\int\limits^{2\pi}_{0}d\phi  \int\limits^{1}_{-1} \mu \f d \mu,
\ee
are given, one can obtain a distribution of radiation in terms of Lagrange multipliers $\eta$ and $a$ \citep[e.g.,][]{smit:00}:
\be
\label{eq:f_eta_a}
\f = \frac{1}{e^{\eta-a\mu}+1},
\ee
where $\mu=\cos\theta$. The second moment of equation (\ref{eq:f_eta_a}) yields $p$ as a function of $\eta$ and $a$. The closure relation is obtained by expressing $\eta$ and $a$ in terms of $e$ and $f$ through inversion of $e(\eta,a)$ and $f(\eta,a)$:
\be\label{eq:p_mefd}
	p=\frac{1}{3}+\frac{2}{3}(1-e)(1-2e) \chi \left(\frac{f}{1-e}\right),
\ee
where $\chi(x) = 1- 3/q(x)$ and $q(x)$ is the inverse Langevin function $L(q)\equiv \coth q - 1/q$. The lowest-order polynomial approximation to function $\chi(x)$ that has the correct free-streaming and diffusive limits is
\be\label{eq:chi_approx}
\chi(x) = x^2(3-x+3x^2)/5,
\ee
which is $\sim 2\%$ accurate \citep{cernohorsky:94,smit:00}. The substitution of this approximation into equation (\ref{eq:p_mefd}) yields an analytical closure that is a function of both $f$ and $e$. We refer to this closure as the {\tt MEFD} closure. It is shown in Fig.~\ref{fig:closures} as a a series of curves for $e=0.1,\ 0.3,\ 0.5,\ 0.7$ and $0.9$ with dashed lines. Note that, in the limit of maximum packing, the {\tt MEFD} closure reduces to \citep[e.g.,][]{smit:00}
\be
	p=\frac{1}{3}\left(1-2f + 4f^2 \right).
\ee
This closure, shown with the bottom curve in Fig.~\ref{fig:closures}, represents one boundary of the {\tt MEFD} closure. The other boundary is the classical limit of this closure, the Maximum Entropy (ME) closure, discussed below.

\subsection{ME: Maximum Entropy Closure in the Classical Limit}

The classical limit of the {\tt MEFD} closure is the closure by \citet{minerbo:78}. It can be obtained from equations (\ref{eq:p_mefd})-(\ref{eq:chi_approx}) by formally taking the $e\rightarrow 0$ limit:
\be
	p = \frac{1}{3} +\frac{2f^2}{15} (3-f +3f^2).
\ee
This closure is shown with the solid sky blue line in Fig.~\ref{fig:closures}. We refer to this closure as the {\tt ME} closure.

\subsection{Janka Closures}

Based on extensive Monte Carlo neutrino transport calculations in PNS envelopes, \citet{janka:91phd,janka:92c} constructed several analytic fits to energy-averaged radiation fields, which were parametrized as 
\be
	p=\frac{1}{3}\left[1+ a f^{m} + (2-a) f^{n} \right],
\ee
where $a$, $n$, and $m$ are the fitting parameters. We consider two closures corresponding to sets $\{a=0.5,b=1.3064,n=4.1342\}$ {\tt Janka\_1} and $\{a=1,b=1.345,n=5.1717\}$ {\tt Janka\_2}. The former is obtained by combining the MC outputs for electron neutrinos from two matter distribution models corresponding to extended hot shocked mantle and compact post bounce configuration. The latter closure is obtained from the $\nu_\mu$ radiation field of the matter configuration at $300\,\mathrm{ms}$ after bounce. These two closures are shown in Fig.~\ref{fig:closures}  with dark and bright purple colors, respectively. 

\section{Results}
\label{sec:results}

In order to asses the quality of M1 results, we consider the radiation field in and around the uniform sphere (Section~\ref{sec:sphere}) and a set of protoneutron star models (Section~\ref{sec:pns}). The former case has an analytical solution, while the latter is calculated with the MC method using the code of \citet{abdikamalov:12}. Both of these problems have the central opaque region and outer transparent envelope
common to many astrophysical sources.

\subsection{Quantitative Estimate of Accuracy}
\label{quant}

In order to estimate the accuracy of the M1 results, we use the normalized mean square deviation and the spectrum-weighted mean square deviation. The former is defined as
\be\label{meansq}
  \d {Y}(X) = \sqrt{\frac{1}{N_X}\sum\limits^{X_{\rm max}}_{X_{\rm min}}
  \left[{1-\frac{Y(X_i)}{Y_{0}(X_i)}}\right]^2}.
\ee
Here, $Y$ stands for any quantity we want to compare (e.g., energy density, flux factor, and etc.), while $Y_0$ is the ``exact'' value of this quantity obtained from the analytical solution or a Monte Carlo calculation. $X$ is a variable on which both $Y$ and $Y_0$ depend (e.g., the radial coordinate) and $X_i$ are its discrete values ranging from $X_{\rm min}$ to $X_{\rm max}$. Thus, $\d {Y}$ provides an estimate of how well the closure solution approximates the exact solution in the entire range from $X_{\rm min}$ to $X_{\rm max}$. 

The spectrum-weighted mean square deviation is defined as
\be\label{spw}
  \bar{\d} {Y} = \frac{\sum w_{i} {\d Y_i}}{\sum w_{i}}, \qquad w_i = S_i/S_{\rm max},
\ee
where $i$ is the index of the neutrino energy group and $\d Y_i$ is defined by equation \rf{meansq} for each energy group independently. The spectral weights $w_i$ are obtained using the spectral energy density $S_i$ at energy $\eps_i$ and the peak value of spectral energy density $S_{\rm max}$. In our analysis of the spectral weighted quantities, we restrict ourselves to the energies lying near the spectral peak. More specifically, we consider only the energy groups with spectral energy densities greater than $0.3S_{\rm max}$ in order to cut out low statistics energy groups.

\subsection{Uniform Sphere}
\label{sec:sphere}

\begin{figure}
\centering
\includegraphics[width=8.7cm]{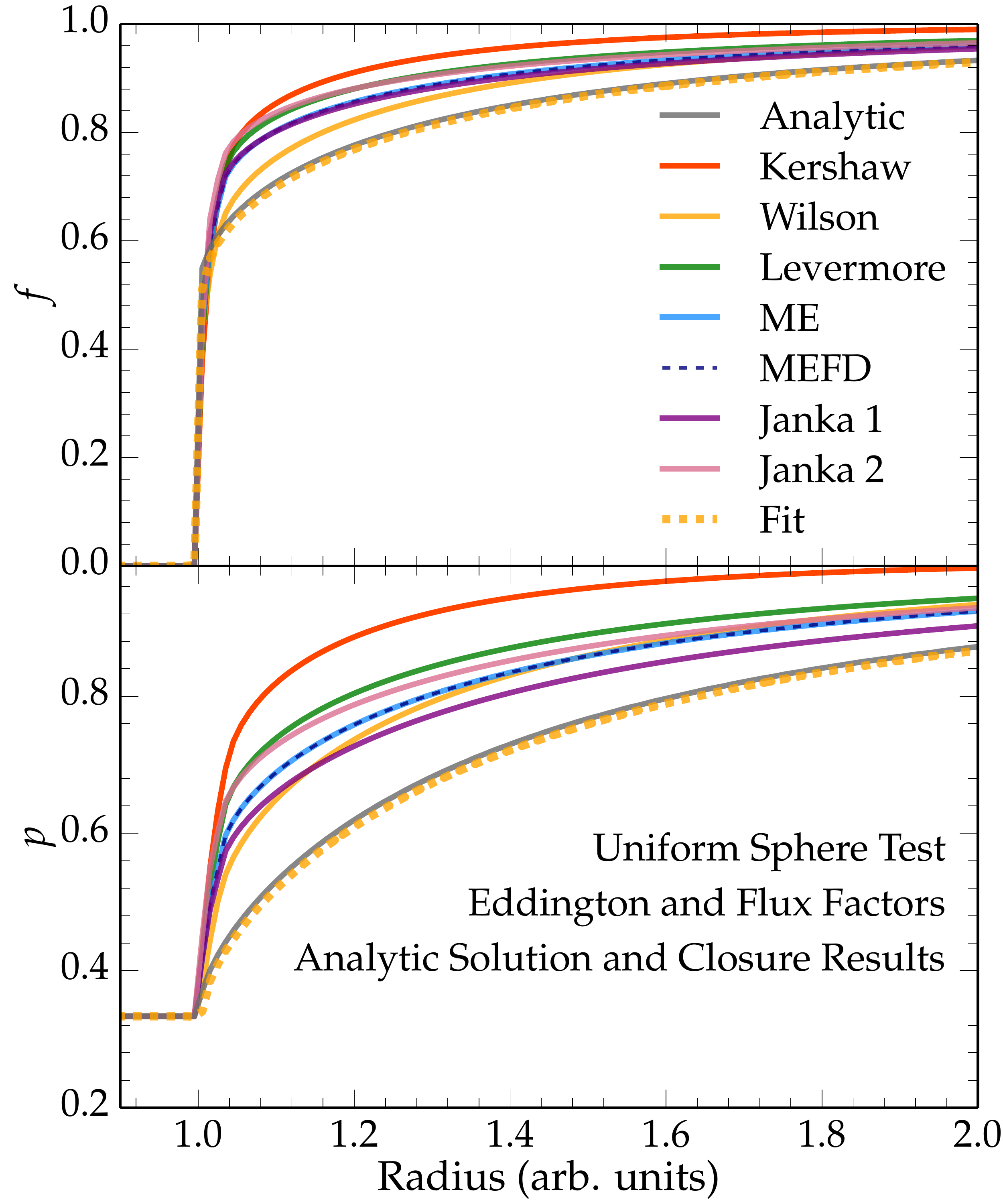}
\caption{The flux factor $f$ and the Eddington factor $p$ as a function of radial coordinate. The matter background is a uniform radiative sphere with $\kappa R = 7500$. The dim gray line is the analytical solution and the colorful lines are the M1 approximations. The performance of the closures is quantitatively evaluated in Table \ref{tab:sph}. The dashed yellow line (Fit) belongs to (\ref{sph_ap}), which is the fit to analytical closure obtained from (\ref{sph_an}).  \label{fig:fp_sph}} 
\includegraphics[width=8.7cm]{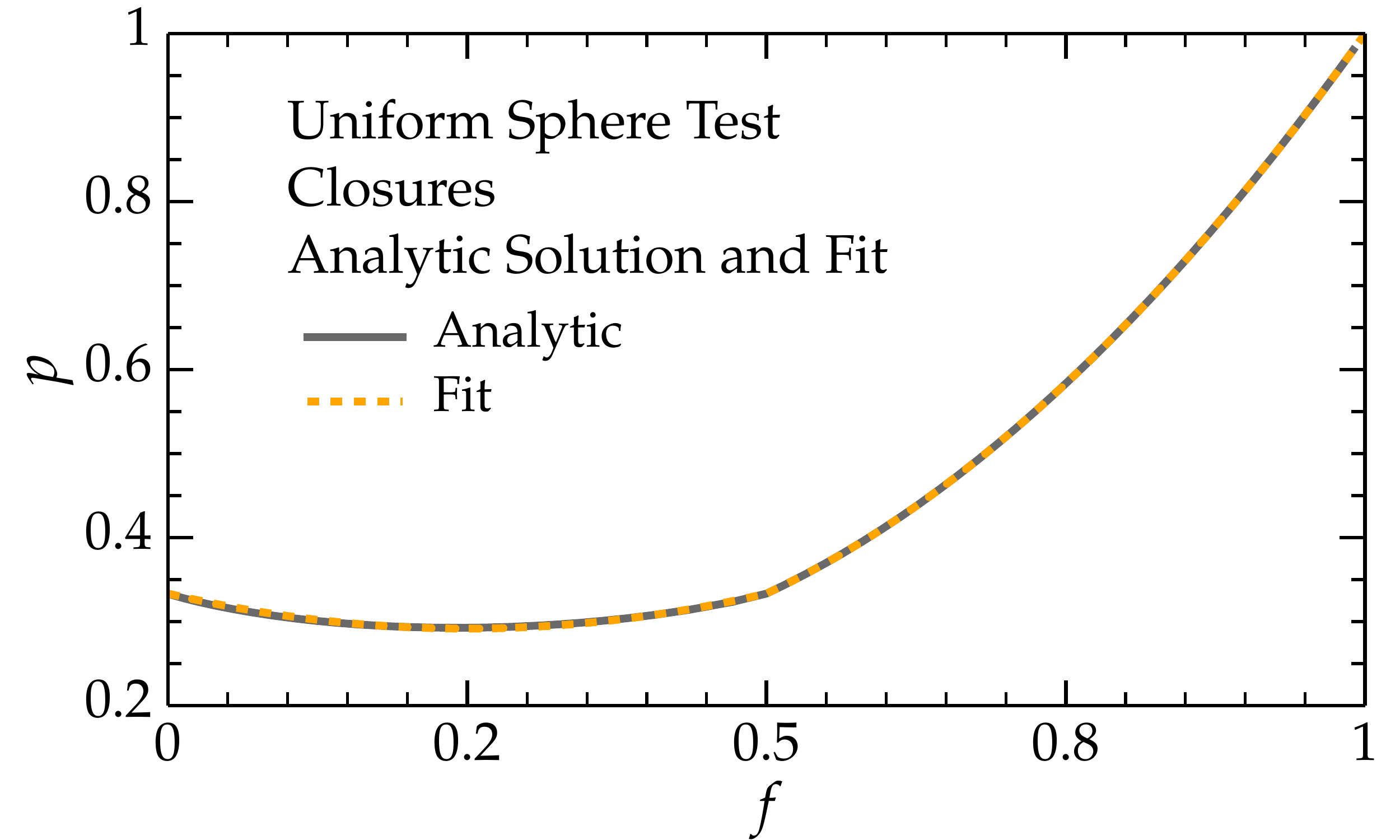}
\caption{The closure reconstructed from the exact solution (\ref{sph_an}) of the uniform radiative sphere problem (solid black line) and analytical fit (\ref{sph_ap}) to that solution (dashed yellow line). This closure is noticeably different from the analytical closures shown in Fig.~\ref{fig:closures}. This explains why these closures yield poor results for the uniform sphere problem. 
\label{fig:sph_closure}} 
\end{figure}

\begin{table}
\begin{center}
  \begin{tabular}{|l|c|c|}
    \hline 
    Closure prescription & $\d f_\nu(r)$ & $\d p_\nu(r)$ \\ \hline
    {\tt Kershaw}   & 0.13 & 0.32 \\
    {\tt Wilson}    & 0.05 & 0.14 \\
    {\tt Levermore}   & 0.10 & 0.22   \\
    {\tt ME}      & 0.07 & 0.17   \\
    {\tt MEFD}    & 0.07 & 0.17   \\
    {\tt Janka\_1}   & 0.07 & 0.13   \\
    {\tt Janka\_2}   & 0.10 & 0.21   \\
   Fit Closure & 0.01 & 0.01 \\ \hline
  \end{tabular} 
  \caption{Mean square deviation of the flux and the Eddington factors obtained with closure prescriptions from the analytical solution for the radiative uniform sphere problem. The sum in the formula for the normalized mean square deviation (\ref{meansq}) is taken over radii from $r_{\rm min} = 1.0$ to $r_{\rm max} = 2.0$.}
\label{tab:sph}
\end{center}
\end{table}

The uniform sphere problem consists of a static homogeneous and isothermal sphere of radius $R$ surrounded by vacuum. Matter inside the sphere can absorb and emit radiation. This problem has an analytical solution and possesses important physical and numerical characteristics. The central opaque source with transparent outer regions are characteristics of many astrophysical systems, while the sharp surface represents a serious challenge for many numerical techniques. For this reason, this problem is often used as a test problem for radiation transport codes~\citep{schinder:89, smit:97, rampp:02,oconnor:15a}

The analytical solution for the distribution function is given as 
\be
\label{sph_an}
\f(r,\mu) = B\left[1-e^{-\kappa R s(r,\mu)}\right] ,
\ee
where $r$ is the radial coordinate, $R$ is the radius of the sphere, $\m = \cos \th$,
\be
\no
s(r,\mu)=\left\{
\begin{array}{lll}
\size \frac{r}{R}\mu+g(r,\mu) & \quad & \mathrm{if} \ \ r<R, \quad -1\le\mu\le 1 \,  \\ 
& \\[-5pt]
\size 2g(r,\mu) & & \mathrm{if} \ \ r\ge R, \quad
\sqrt{1-\left(\frac{R}{r}\right)^2} \le\mu\le 1 \, \\ & \\ 
\size 0 & & \mathrm{otherwise} 
\end{array}
\right.
\ee
and
\be
\label{eq:an_sol_g}
g(r,\mu) = \sqrt{1-\left(\frac{r}{R}\right)^2 (1-\mu^2)} \, .
\ee
Inside the sphere the absorption coefficient $\kappa$ and emissivity $B$ are constants. Outside the sphere, there is no emission and absorption. For our test, we use $\kappa R = 7500$ and $B=1$, which ensures that radiation is fully isotropic inside the sphere and a tiny region $\sim 1/\kappa\ll R$ separates it from the free-streaming regime outside the sphere. 

The flux factor $f$ (top panel) and the Eddington factor $p$ (bottom panel) as a function of the radial coordinate are shown on Fig. \ref{fig:fp_sph}. The dim gray line represents the analytical solution, while the rest of the lines represent the solutions obtained with M1 approximations. The values of normalized mean square deviations of these solutions from the analytical result are listed in Table \ref{tab:sph}. 

As we see, all closures perform poorly for this problem. The {\tt Kershaw} closure yields significantly worse results than the rest of the closures. The normalized mean square deviation of the flux and Eddington factors obtained with closure prescriptions from analytical result are $0.13$ and $0.32$, respectively (cf. Table~\ref{tab:sph}). The next worse performers are the {\tt Levermore} and {\tt Janka\_2} closures. The normalized deviations of the Eddington factor are $0.22$ and $0.21$ for these two closures, respectively. The best performers are the {\tt Wilson} and the {\tt Janka\_1} closures, for which the normalized deviations of the Eddington factor are $0.14$ and $0.13$. The rest of the closures yield intermediate results. 

Such a poor result of the analytical closures has a simple explanation. The true closure for this problem, which can be reconstructed directly from the solution \rf{sph_an}, is shown with a solid black line in Fig.~\ref{fig:sph_closure}. This closure is significantly different from all the aforementioned closures, as we can glean by comparing Fig.~\ref{fig:sph_closure} with Fig.~\ref{fig:fp_sph}. Therefore, the reason why our closures yield inaccurate results is simply because these closures are different from the true closure for this problem. 

Interestingly, the true closure in Fig.~\ref{fig:sph_closure} can be fit well with the following simple analytical expression
\be\label{sph_ap}
	p = \left\{
	\ba{ll}
		1/3-1/3f + 2/3 f^2, & \qquad f \leq 1/2,\\
		1/3-2/3f + 4/3 f^2, & \qquad  f>1/2,
	\ea
	\right. 
\ee
which is shown with the orange dashed line in Fig. \ref{fig:sph_closure}. If we perform M1 calculations with this closure, it reproduces the analytical result (\ref{sph_an}) with excellent $\sim 1 \%$ accuracy (cf. Table \ref{tab:sph}). Note that this fit closure is not expected to perform well for any other matter distributions except the uniform sphere. It is specific to this particular model problem.

\subsection{Protoneutron Star}
\label{sec:pns}

\begin{figure}
  \centering
  \includegraphics[width=8.7cm]{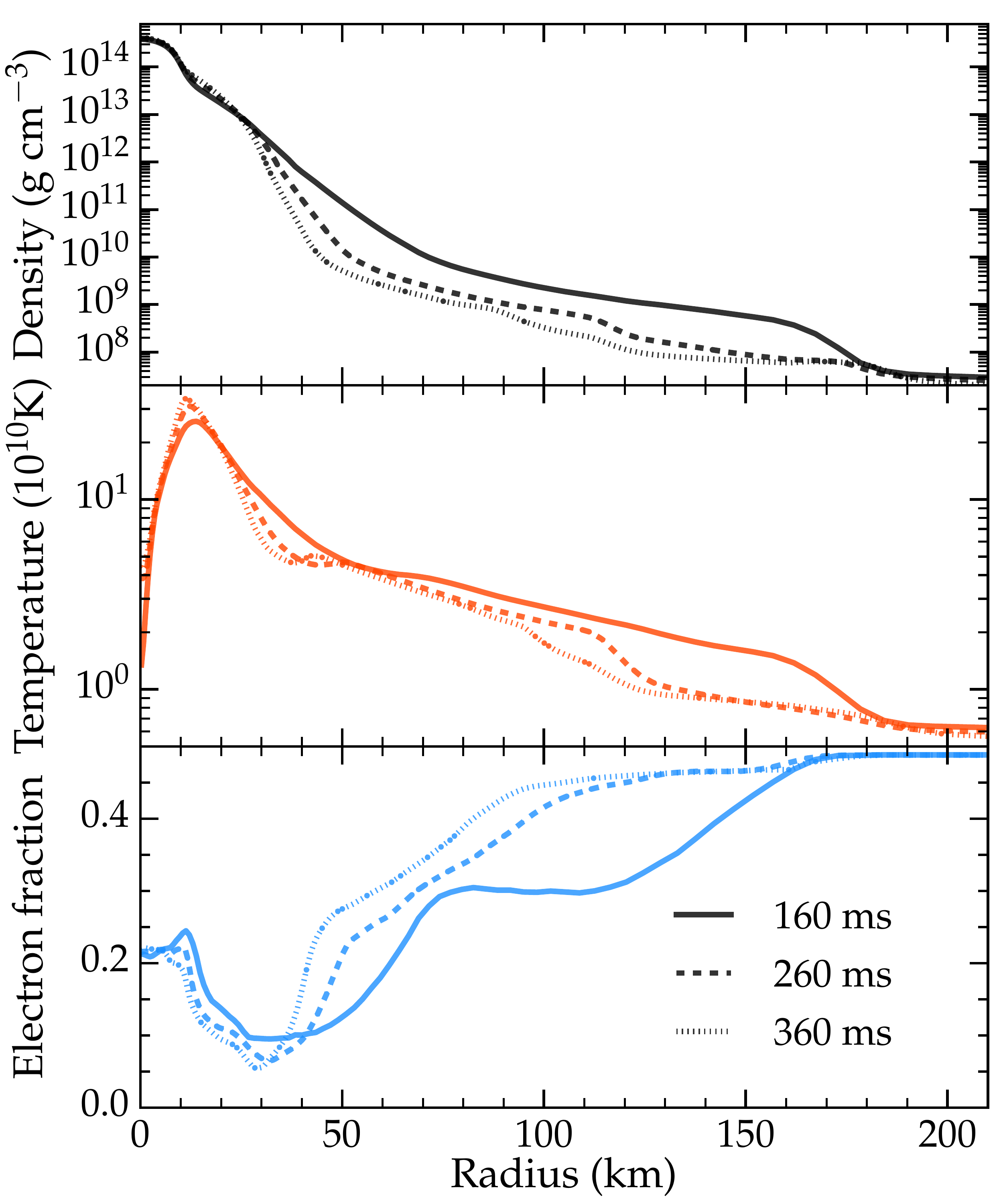}
  \caption{The radial profiles of density (upper pane), temperature (center panel), and electron fraction (bottom panel) for protoneutron star models of \citet{ott:2008} at 160 ms (solid line), 260 ms (dashed line) and 360 ms (dash-dotted line) after bounce. The radial profiles are obtained by angular averaging the 2D data of \citet{ott:2008}.\label{fig:pns}}
\end{figure}

\begin{figure}
  \centering
  	\includegraphics[width=0.95\columnwidth]{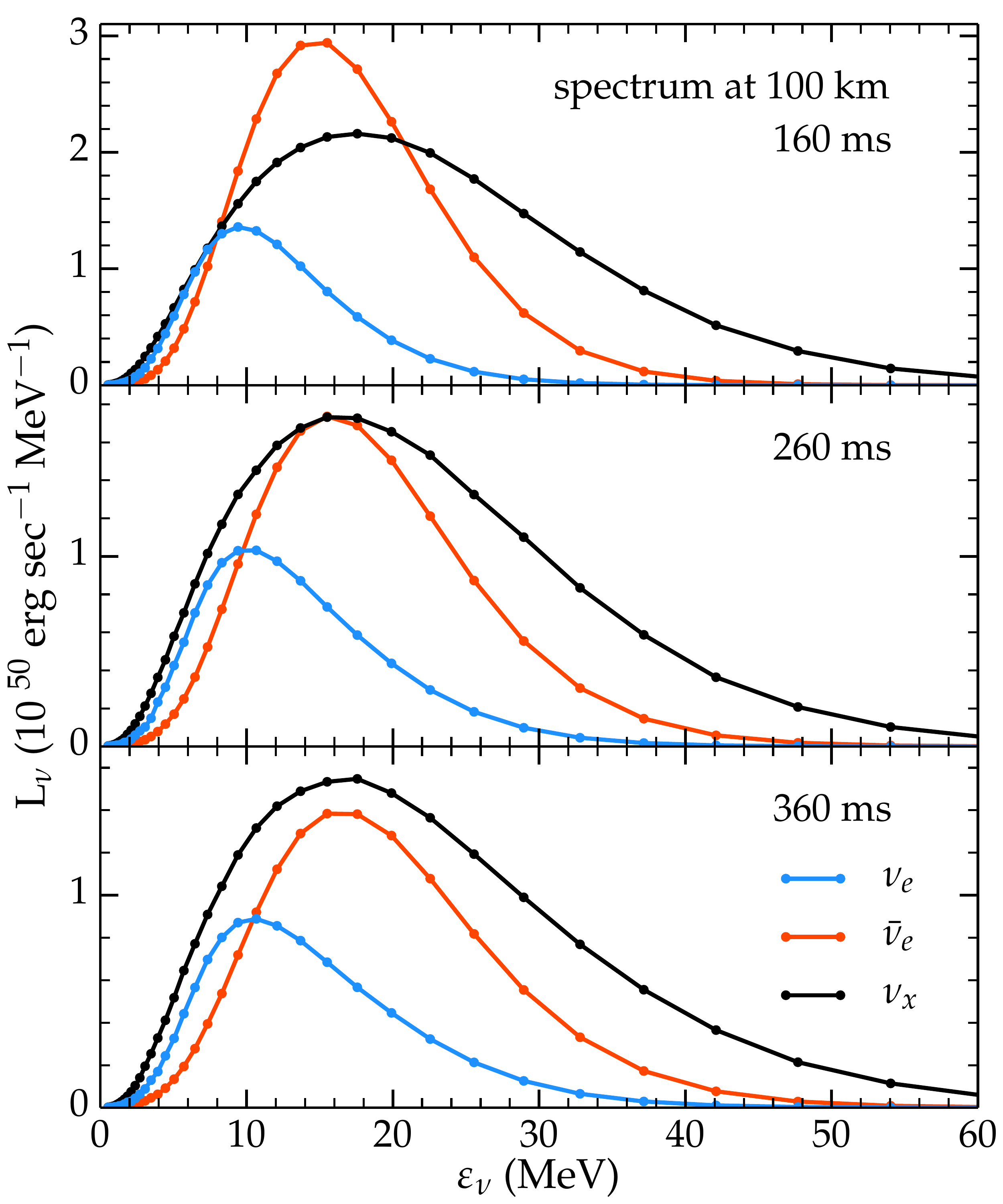}
  \caption{The spectra of neutrino luminosity measured at a radius of $100\,\mathrm{km}$ obtained from MC code. The {\bf top}, {\bf center}, and {\bf bottom panels} represent the PNS models of \citet{ott:2008} at 160, 260, and 360 ms after bounce, respectively. \label{fig:pns_lum}}
\end{figure}

\begin{figure*}
  \centering
  \includegraphics[width=2\columnwidth]{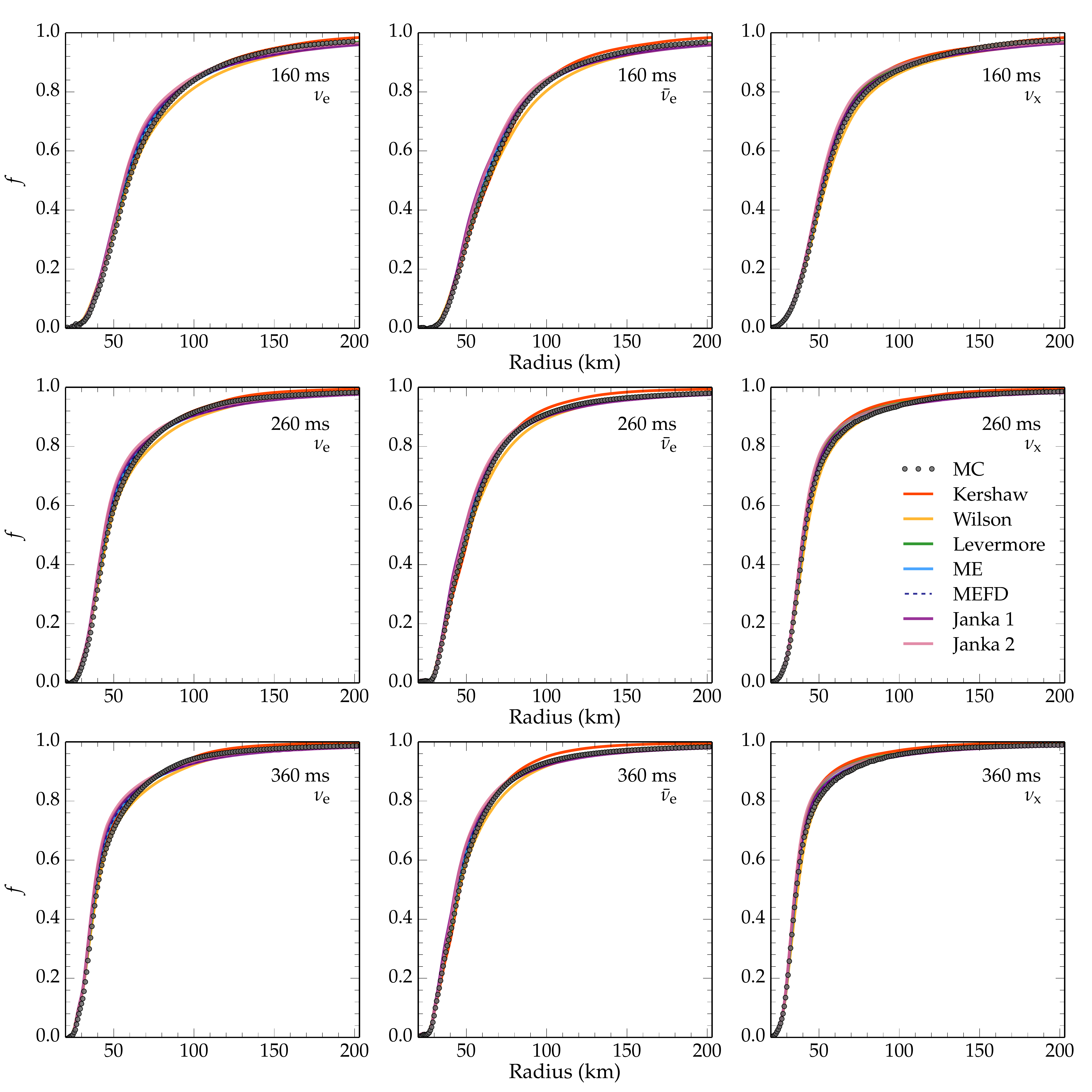}
  \caption{The radial profiles of the flux factor $f$ obtained using the seven closures and from the Monte Carlo code (dotted line) for three different types neutrinos and three PNS configurations at 160, 260, and 360 ms after bounce. All quantities are measured at the neutrino energy groups corresponding to the peak luminosity at $100\,\mathrm{km}$ for each neutrino type. The spectra of neutrino luminosities at this radius are shown in Fig.~\ref{fig:pns_lum} for each neutrino type and three PNS models.\label{fig:pns2}}
\end{figure*}

\begin{figure*}
  \centering
  \includegraphics[width=0.8\columnwidth,clip=false]{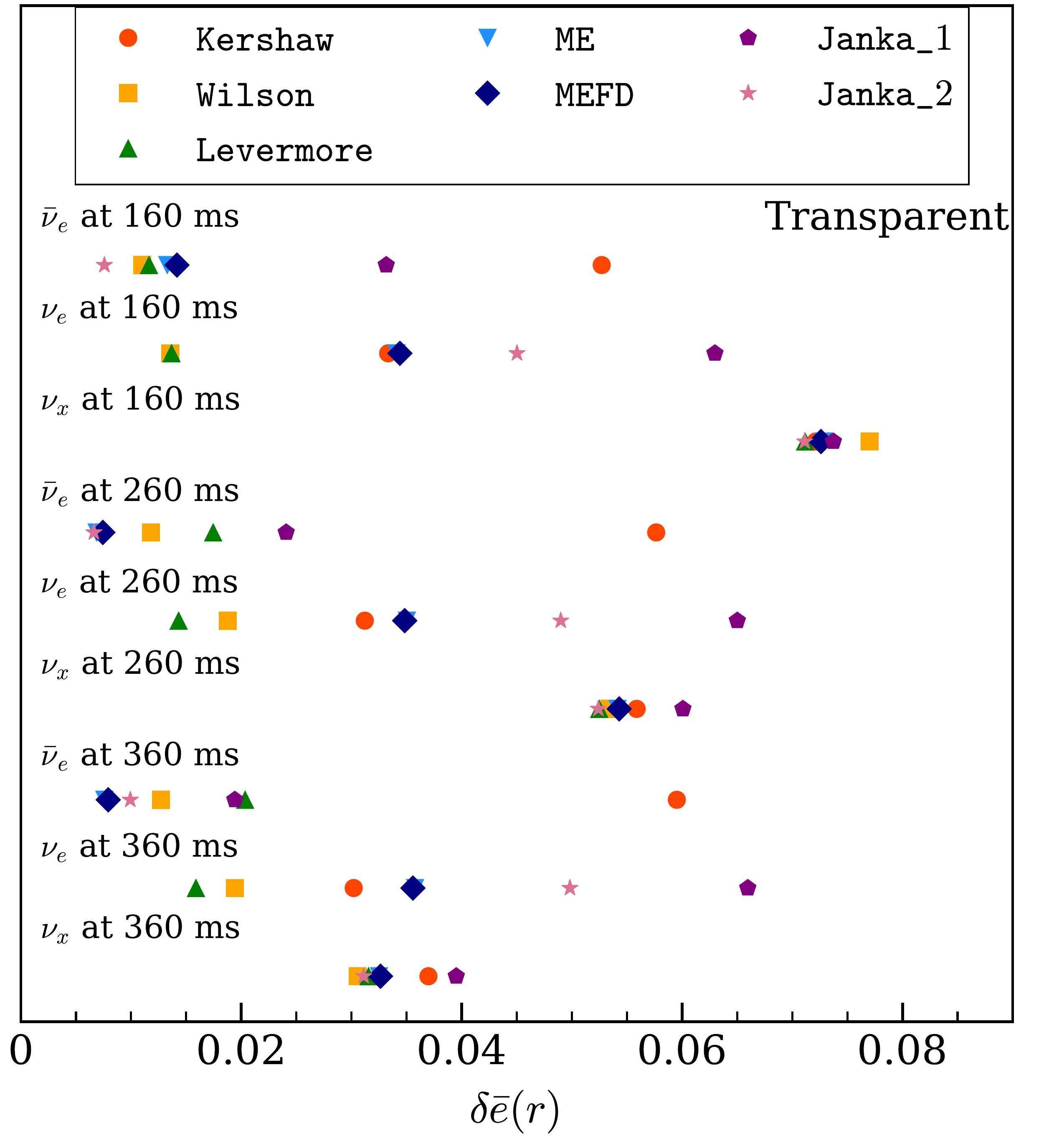}
  \includegraphics[width=0.8\columnwidth,clip=false]{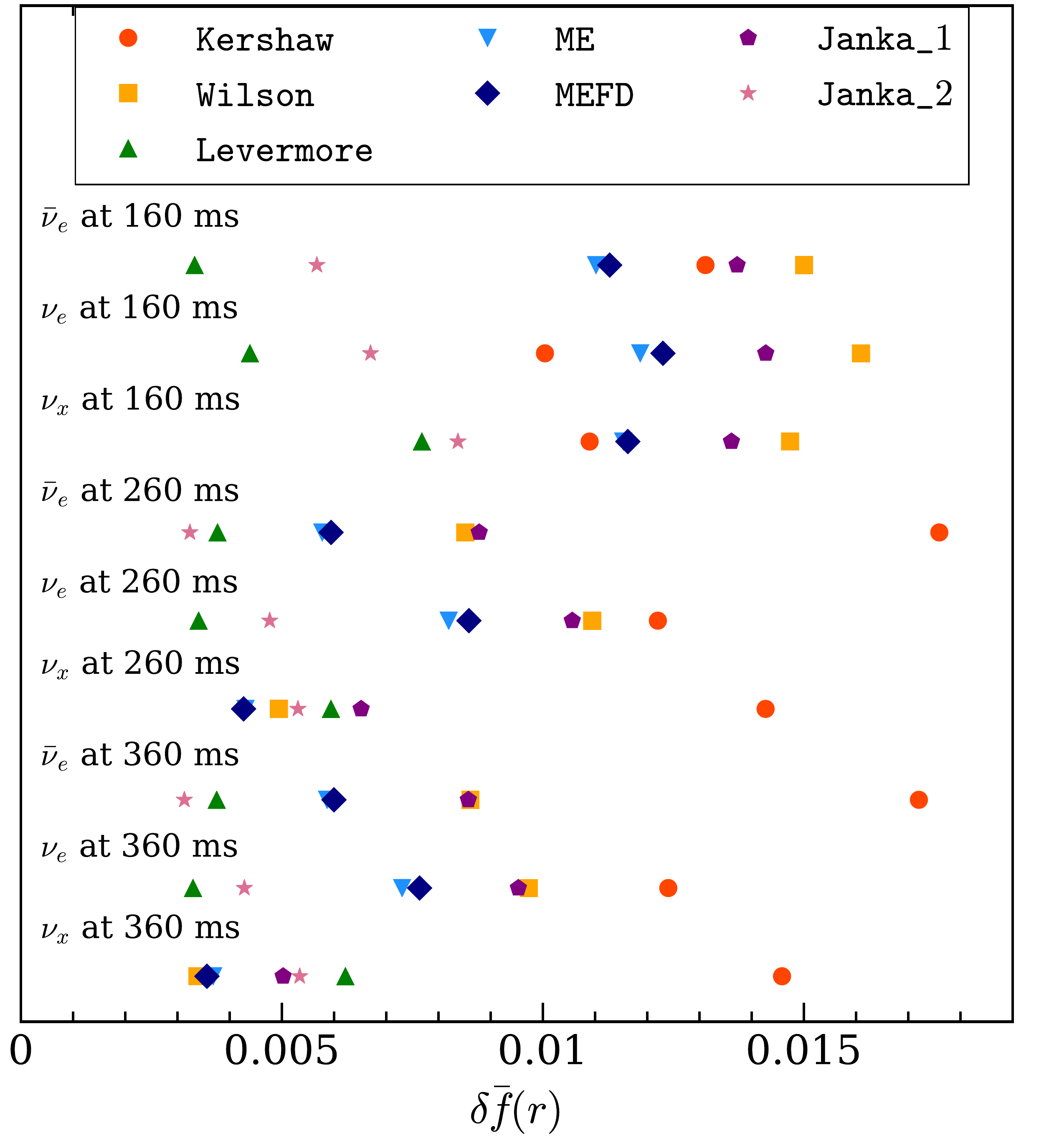}
  \includegraphics[width=0.8\columnwidth,clip=false]{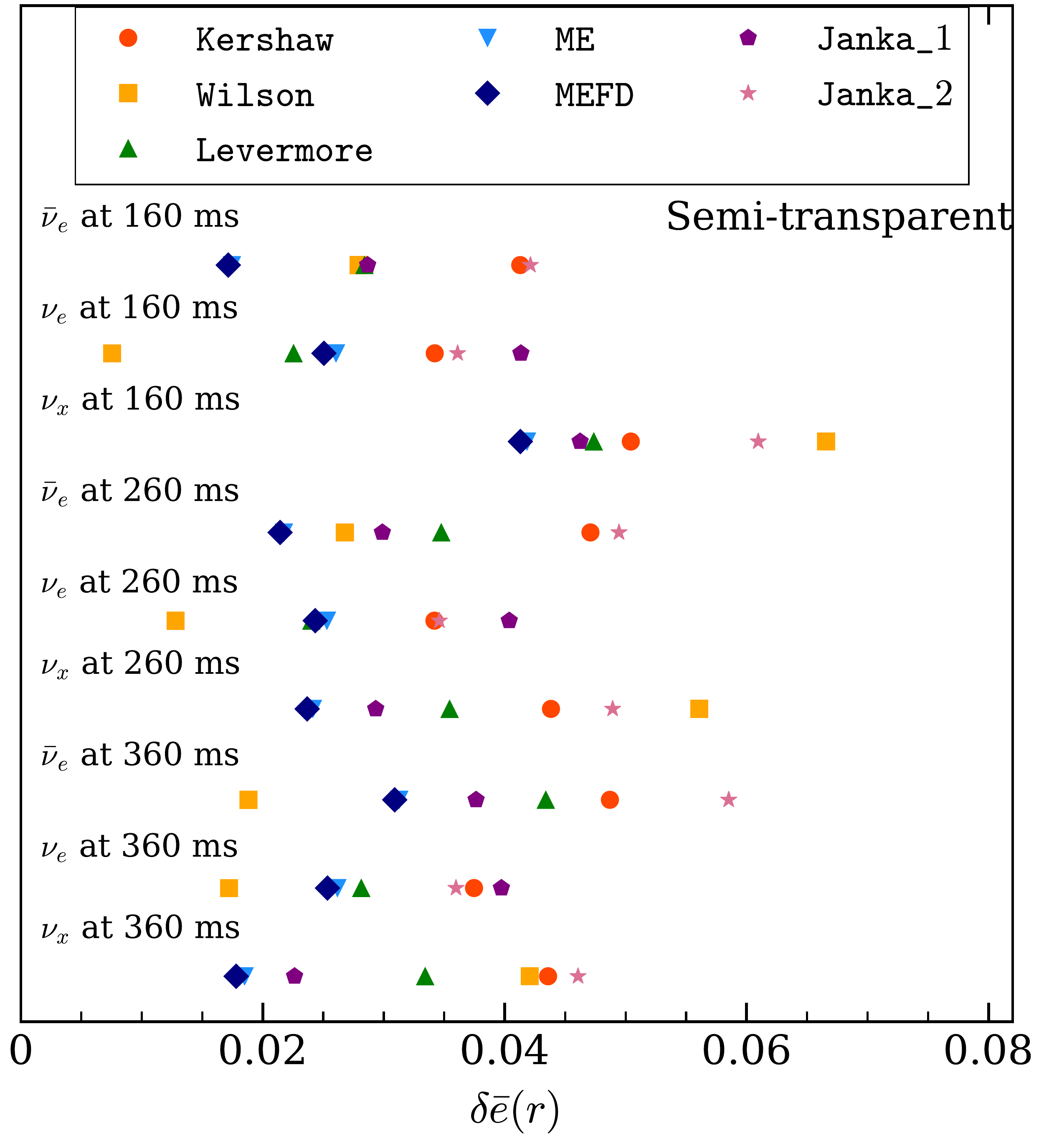}
  \includegraphics[width=0.8\columnwidth,clip=false]{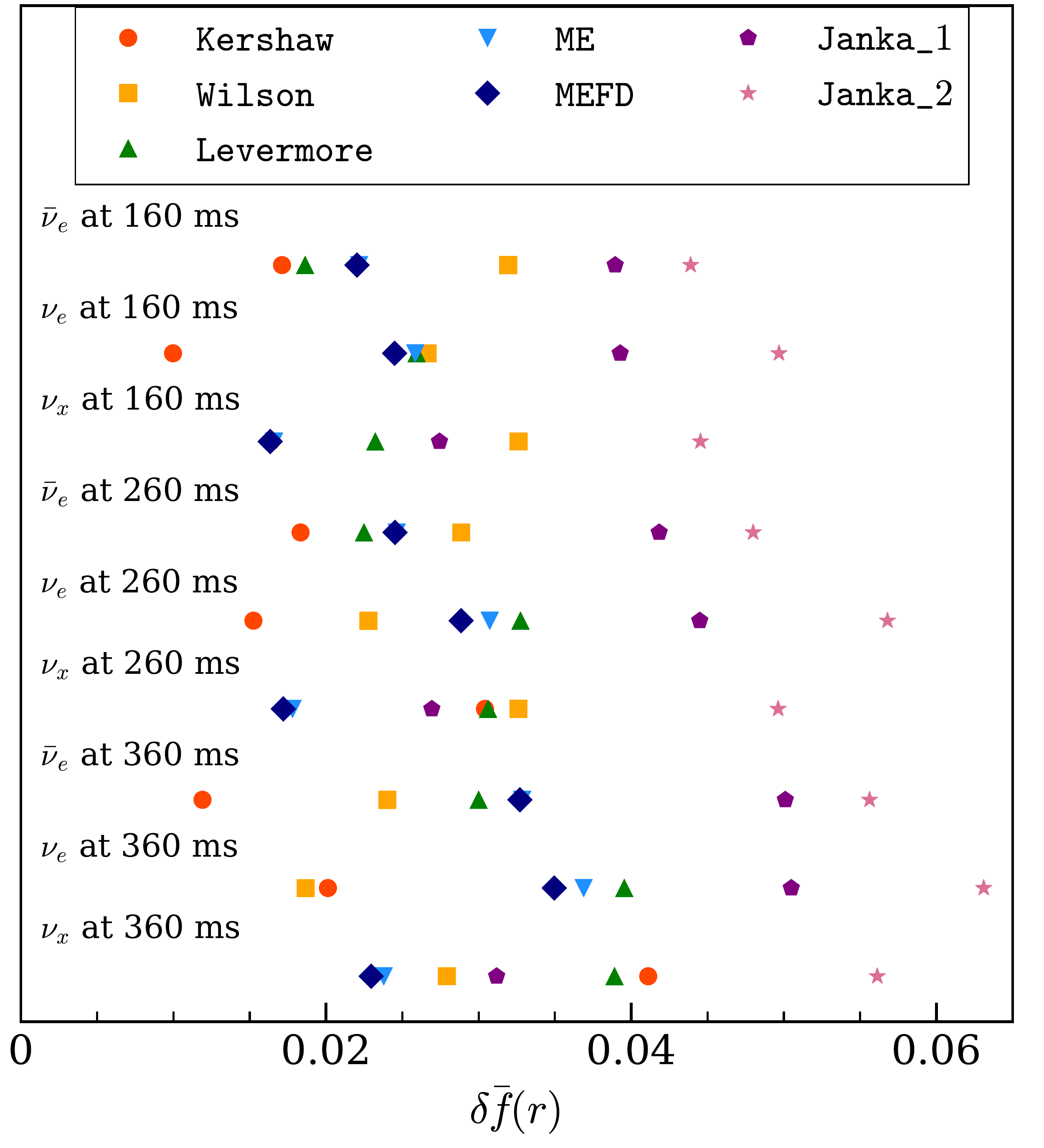}
  \includegraphics[width=0.8\columnwidth,clip=false]{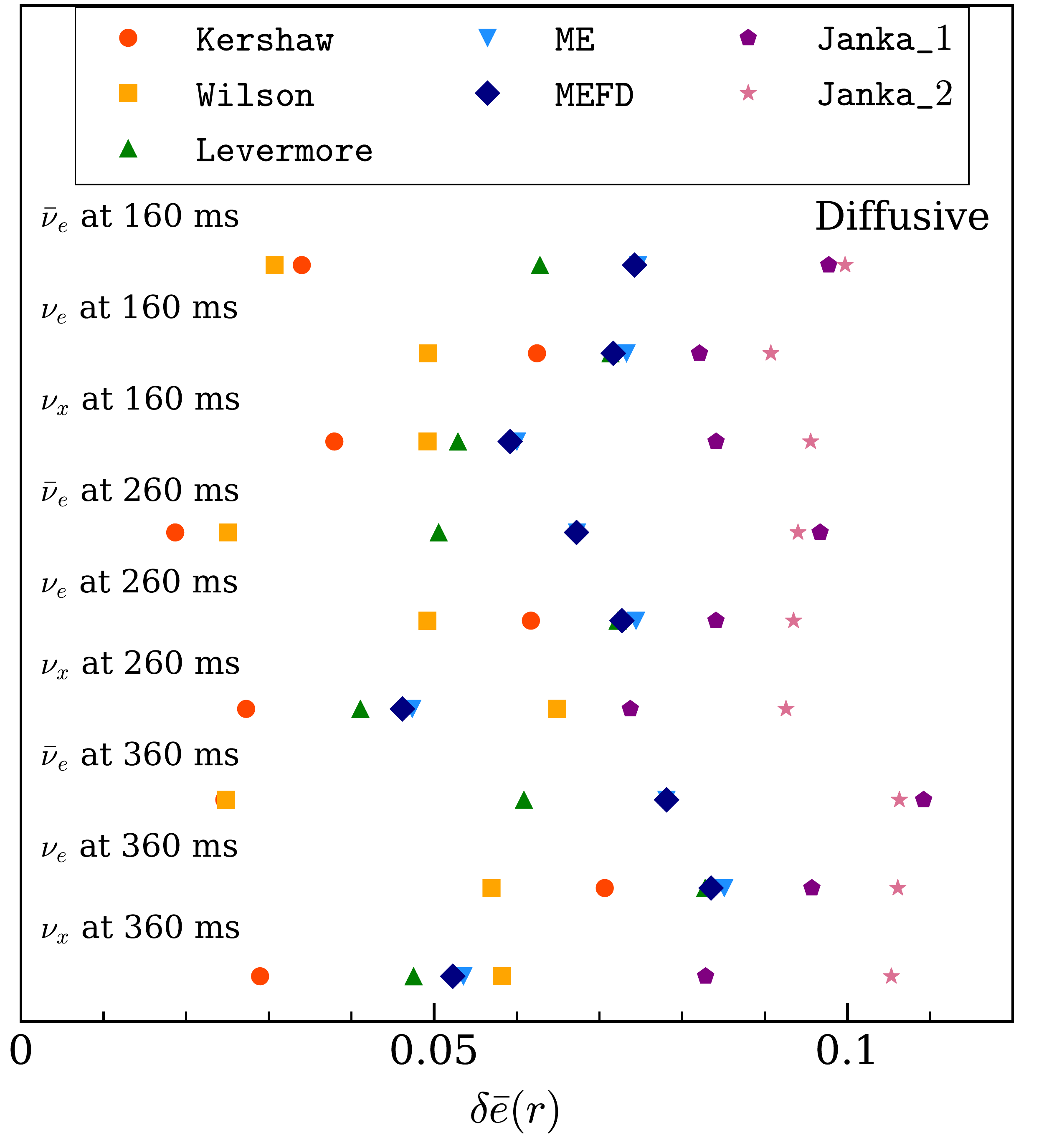}
  \includegraphics[width=0.8\columnwidth,clip=false]{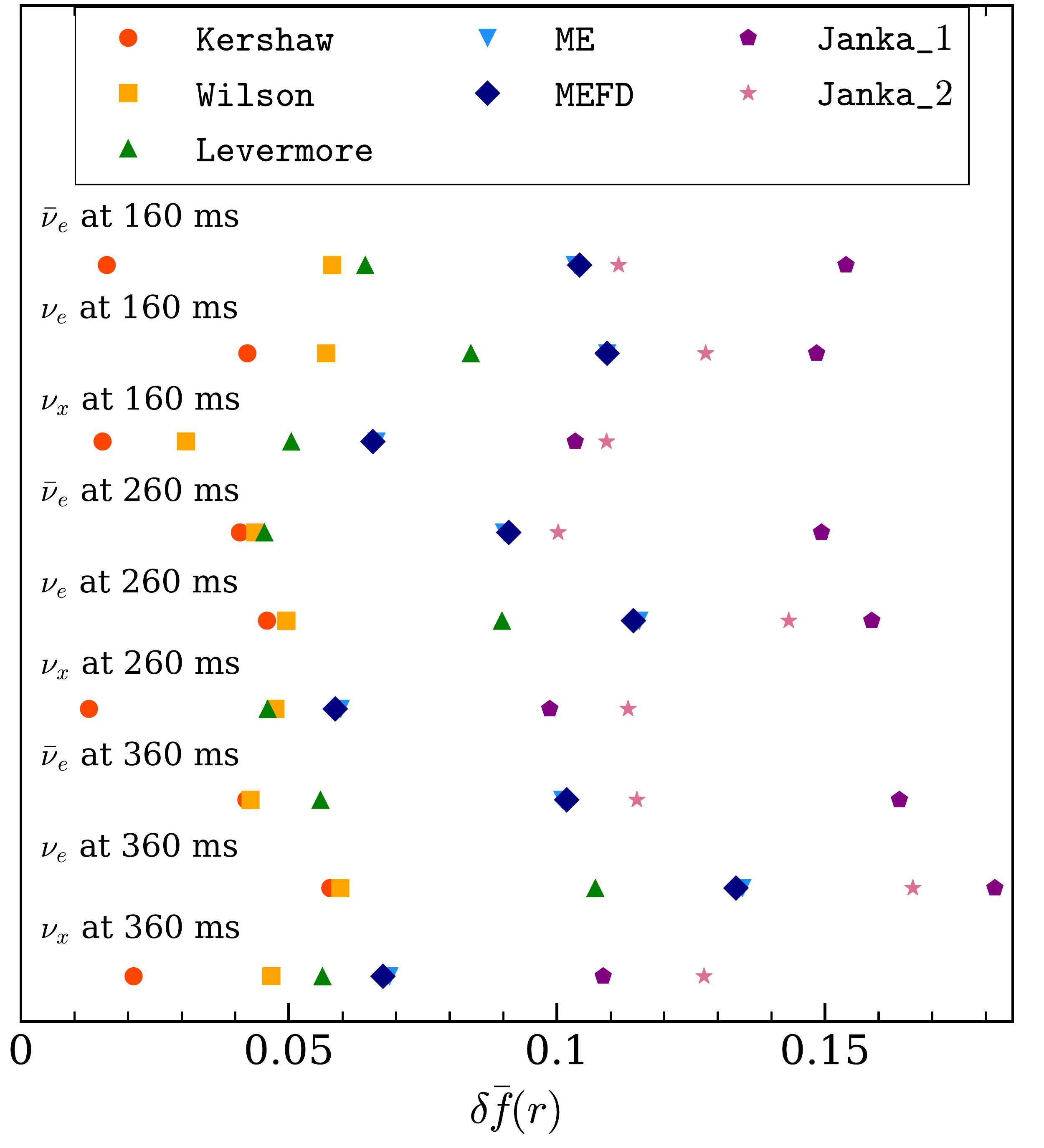}
  \caption{The spectrum-weighed mean square deviation of energy density $e$ (left panel) and the flux factor $f$ (right panel) for different neutrino types for PNS models at $160$, $260$, and $360$ ms after bounce. The top, center, and bottom panels represent the transparent ($0.9 \le f \le 1$), semi-transparent ($0.5 \le f \le 0.9$), and diffusive ($0.2 \le f \le 0.5$) regimes, respectively.    
  \label{fig:ef_dev}}
\end{figure*} 

In this section, we evaluate the ability of the M1 closures to model the neutrino radiation field around spherically symmetric models of protoneutron stars (PNSs) formed in core-collapse supernovae. We take three different post-bounce configurations
(obtained from 2D radiation-hydrodynamics simulations of \citealt{ott:2008}) of a $20M_\odot$ progenitor star at 160 ms, 260 ms, and 360 ms after bounce. We average the 2D profiles of \citet{ott:2008} over angle to obtain spherically symmetric configurations of PNSs. The radial profiles of density, temperature, and electron fraction are shown in upper, center, and bottom panels of Fig.~\ref{fig:pns}. The spectra of neutrino luminosity obtained from MC code at the radius of $100\,\mathrm{km}$ are shown in Fig.~\ref{fig:pns_lum}. The top, center, and bottom panels represent the PNS models at 160, 260, and 360 ms after bounce, respectively.

The ``exact'' solution of the problem is obtained from MC calculations using the code of \cite{abdikamalov:12}, while the M1 results are obtained using the {\tt GR1D} code \citep{oconnor:15a}. We evolve our time dependent MC code until we reach steady-state neutrino radiation field for each model of PNS. This field is then averaged over many timesteps until we get rid of the stochasticity in the MC solution. Since the MC method does not use any approximations in the solution procedure \citep{abdikamalov:12}, the solution obtained this way is exact for a given configuration of matter (i.e., a given configuration of opacities and emissivities). In order to ensure the consistency of the results, the two codes use identical microphysical inputs. Both use the \citet{shen:98a} equation of state (EOS) table and a {\tt NuLib} opacity table \citep{oconnor:15a} that was generated with the same EOS table. In both codes, we use 48 logarithmic energy groups ranging from $0.5\,\mathrm{MeV}$ to $200\,\mathrm{MeV}$. In the MC code, we use $150$ radial logarithmically spaced zones with the central resolution of $0.2\,\mathrm{km}$. We have performed extensive resolution tests in order to ensure that our results are convergent. 

We examine the quality of the closures in the free streaming, semi-transparent, and opaque regimes. We separate these regimes based on the value of the flux factor $f$. We choose the transparent regime as the one where $0.9<f<1$, the  semi-transparent as $0.5 \le f \le 0.9$ and the opaque as $0.2 \le f \le 0.5$. We neglect the region of low $f$ because the MC results suffer from noise in the highly diffusive region. It is more appropriate to separate the different regimes based on $f$ rather than, e.g., the value of the radial coordinate, because at a given radius, neutrinos of different energies behave differently. For example,  low-energy neutrinos have lower opacity and hence they propagate more freely compared to higher-energy neutrinos at the same radius.

The radial profiles of the flux factor $f$ obtained using the seven closures and from the Monte Carlo code (dotted line) for three different types of neutrinos and three PNS configurations at 160, 260, and 360 ms after bounce are shown in Fig.~\ref{fig:pns2}. All flux factors are measured at the neutrino energy groups corresponding to the peak luminosity at $100\,\mathrm{km}$ for each neutrino type (cf. Fig.~\ref{fig:pns_lum}). As we can see, all M1 closures yield qualitatively correct results. 

For a more precise quantitative estimate, we utilize the spectrum-weighed deviations of the flux factor and energy density from the MC results, equation \ref{spw}. The spectrum-weighted deviation $\bar\delta e$ of the energy density $e$ in the transparent regime for the seven different closures, for the three neutrino types, and the PNS models at $160$, $260$, and $360$ ms are shown on the top left panel of Fig.~\ref{fig:ef_dev}. The deviations $\bar\delta e$ are calculated using formula (\ref{spw}), in which the spectrum is taken at $100\,\mathrm{km}$. In order to verify that our results are not too dependent on spectral information at different locations, we have calculated $\bar\delta e$ using spectra at several different radii and obtain results very similar to $\bar\delta e$ that use spectra at $100\,\mathrm{km}$. This suggests that the values of $\bar\delta e$ presented in Fig.~\ref{fig:ef_dev} are robust measures of errors of the M1 closures for the radiation field around PNSs.

As we can see, different closures yield different levels of accuracy depending on the neutrino type and PNS model. No single closure performs consistently better (or worse) than other closures in all cases. That said, the {\tt Wilson} and the {\tt Levermore} closures perform better than others in most cases in the transparent regime, followed by the {\tt ME} and {\tt MEFD} closures. The {\tt Janka\_1} and the {\tt Janka\_2} closures exhibit $\bar\delta e \gtrsim 0.04$ in most cases, which is worse than $\bar\delta e$ for the rest of the closures. This is a remarkable result because the {\tt Janka\_1} and {\tt Janka\_2} closures were constructed from fitting to the exact solution for the neutrino radiation field around PNS. This demonstrates that a closure constructed for one PNS model (with a given EOS and opacity table) does not necessarily yield a good result for all other PNS models. 

The behavior of the deviation of the flux factor $\bar\delta f$ is shown on the top right panel of Fig.~\ref{fig:ef_dev}. In this case, the {\tt Levermore} closure performs better than other closures in almost all cases and yields small $\bar\delta f \sim 0.004$. The {\tt Janka\_2} closure performs only slightly worse than this closure, yielding $0.004 \lesssim \bar\delta f \lesssim 0.009$. However, this is $\sim 2-3$ times smaller than what the rest of the closures yield, which is a surprising result because the {\tt Janka\_2} closure yields relatively poor result for the energy density compared to most of the closures, as we discussed in the previous paragraph. 

The {\tt Kershaw} closure produces the least accurate $f$ compared to the other closures in all models except the PNS models at $160\,\mathrm{ms}$ after bounce, yielding deviations of $\bar \delta f \sim 0.015$ in all cases. The {\tt Wilson} closure yields intermediate results in most situations except for the PNS model at $160$ ms after bounce, for which it yields the largest deviation of $\bar\delta f \sim 0.015$. This again shows that a closure that yields the best result for the energy density does not necessarily yield the best result for the flux factor. As we see below, the reverse of this statement is also true.  

The spectrum-weighed deviations $\bar\delta e$ and $\bar\delta f$ for the semi-transparent regime are shown in the left and right center panels of Fig.~\ref{fig:ef_dev}, respectively. In this regime, both the {\tt ME} and {\tt MEFD} closures often -- but not always -- yield the smallest deviations $\bar\delta e$. The {\tt Janka\_1} and {\tt Janka\_2} closures yield the largest $\bar\delta f$ of $\sim 0.04-0.06$ in all cases. The {\tt Wilson} closure yields the smallest $\bar\delta e$ in most cases, but produces $\bar \delta f \sim 0.03$, which is roughly the mean of the values of $\bar\delta f$ produced by all of the closures. On the other hand, the {\tt Kershaw} closure yields the smallest $\bar\delta f$ of $\sim 0.01-0.02$ in most cases, but yield relatively large $\bar\delta e$ of $\sim 0.035-0.05$.

In the opaque region, the situation is significantly different. The {\tt Kershaw} closure, which often yields the largest $\bar \delta e$ and $\bar \delta f$ in the transparent and $\bar\delta e$ in the semi-transparent regimes, produces the smallest $\bar \delta e$ and $\bar \delta f$ of $\lesssim 0.05$ in the opaque regime. The {\tt Wilson} closure yields slightly worse results than the {\tt Kershaw} closure with $\bar\delta e \sim \bar\delta f \sim 0.05$. The {\tt Janka\_1} and {\tt Janka\_2} closures yield the largest errors ($\bar \delta e \sim 0.1$ and $\bar \delta f \sim 0.1-0.15$). This again shows that a closure that yields good results for one model of PNS does not necessarily produce good results for other PNS models. 

\begin{figure}
  \centering
  \includegraphics[width=8.7cm]{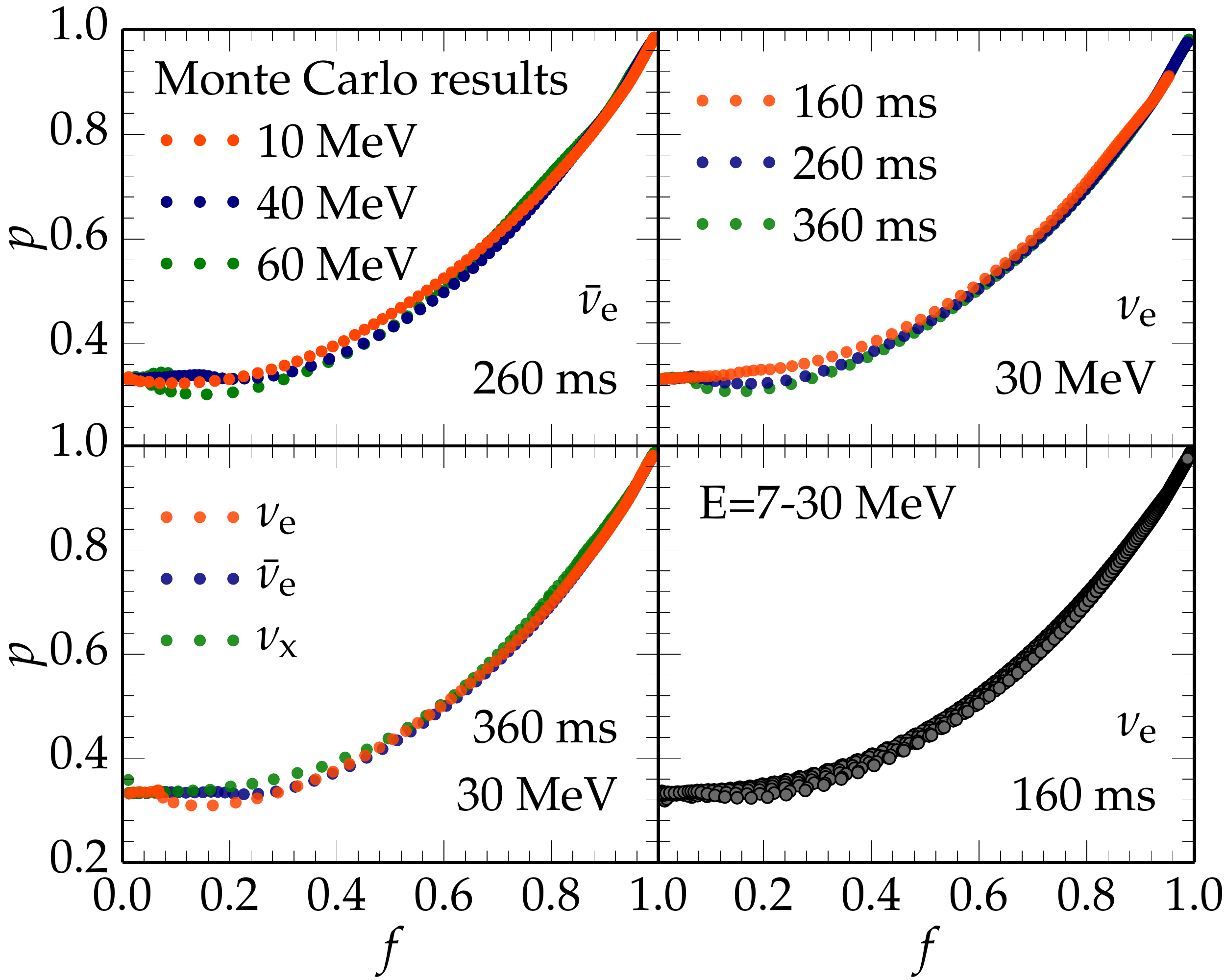} 
  \caption{The closure relations obtained from Monte Carlo results as the function of the neutrino type (top left), the neutrino energy (top right), and the background matter distribution (bottom left). The bottom right shows the variation in the closures as a function of neutrino energy.  Here, twelve groups are plotted together.
  \label{fig:dif_pns}}
\end{figure} 

In all cases, the {\tt ME} and {\tt MEFD} maximum entropy closures yield almost identical results. This is direct consequence of the fact the former closure is the classical limit of the latter and the radiation field, in the regions we considered, is predominately in the classical regime.

Overall, these two closures yield relatively good results in all cases and never result in the largest deviations compared to the other closures. This, in combination with the fact that the {\tt ME} closure is simpler and requires fewer operations to compute than the {\tt MEFD} closure, makes the former a more attractive option for neutrino transport applications involving protoneutron stars. 

These results demonstrate that no single closure performs the best or the worst in all cases. Whether a specific closure is ``correct'' for a given problem depends on the parameters of the problem, such as the matter configuration (i.e., profiles of density, temperature, and composition) and the neutrino type. We find a similar level of differences between deviations of $e$ and $f$ for different neutrino energy groups (not shown here). Fig.~\ref{fig:dif_pns} shows the functional form of the closures $p(f)$ extracted from the MC simulations for different neutrino types, different PNS models and different energy groups. We can clearly see variations between closures $p(f)$ corresponding to different problems. 

\section{Conclusion}
\label{sec:conclusion}

We conducted a systematic, quantitative study of the accuracy of analytical closure relations for two-moment neutrino radiation transport schemes commonly used in the literature. We considered the neutrino field around two sets of radiating objects: the uniform radiative sphere and PNS models at $160$, $260$, and $360$ ms after core bounce obtained from simulations of \cite{ott:2008}. In all cases, the matter configuration is assumed to be static. This restriction allows us to focus on the quality of the closure relations and exclude other sources of errors such as those stemming from non-linear radiation-matter coupling. 

We considered seven different closures. These are the closures by \citet{kershaw:76}, \citet{wilson:75}, \cite{levermore:84}, and the maximum entropy closures of \citet{cernohorsky:94} and \citet{minerbo:78}. In addition, we considered two closures that are constructed by fitting to exact Monte Carlo solutions of the radiation field around PNSs by \citet{janka:91phd}.  

We find that no single closure, among those studied here, is consistently better or worse than any other. A closure that yields accurate results in one case may not yield as good results in other situations. The level of accuracy that a given closure yields varies for different quantities. 

Given this limitation of the closures, the maximum entropy closure by \citet{minerbo:78} and \citet{cernohorsky:94}, which yield almost identical results, often yield better results among all the closures studied. These two closures never yielded the worst results compared to all other closures. In this sense, these two are a safe choice, as they are less likely to yield extremely erroneous result over a wide range of problems and variables. Since the \citet{minerbo:78} closure is simpler to compute than the one by \citet{cernohorsky:94}, we conclude that the former closure is the most attractive choice for problems involving neutrino transport around PNSs.

In this work, we assumed spherical symmetry, static matter, and flat spacetime. These assumptions limit the scope of the implications of our results. In particular, the hydrodynamic and general relativistic effects introduce shifts in the energy spectrum of neutrinos, which alter the moment equations \citep[e.g,][]{cardall:13rad,just:15b}. Also, in non-spherically-symmetric cases, we can have radiation sources at different spatial locations. Notable examples are the accretion disks formed in neutron star mergers and hot spots near the PNS surfaces in the context of core-collapse supernovae. The interaction of radiation beams from such sources cannot be modeled within our spherically-symmetric setup. It is \emph{a priori} unclear the extent to which our results are valid in such cases. This will the subject of a future study.

\section*{Acknowledgements}

We thank Evan P. O'Connor, Christian D. Ott and John Wendell for their input. We also thank Adam Burrows, David Radice, Sherwood Richers, and Luke Roberts for helpful discussions. This work was supported by IGPPS program at Los Alamos National Laboratory (2012-2015) and partially by the Sherman Fairchild Foundation. EM is grateful to David and Barbara Groce for their kindness and support. EA acknowledges support from NU ORAU and Social Policy grants. LANL report number LA-UR-17-20390.

\appendix

\section{Methods}
\label{sec:methods}

\subsection{The \code{GR1D} Radiation-Hydrodynamics Code}

We employ the M1 radiation transport solver that is part of the \code{GR1D} radiation-hydrodynamics Code \ci{oconnor:11,oconnor:13,oconnor:15a} available at \url{http://www.GR1Dcode.org}. The matter in our calculations is static and the metric is Minkowski. There is no coupling between different energy groups. For these conditions, \code{GR1D} implements M1 closures by setting the pressure tensor according to equation (\ref{eq:closure}). Under these conditions, the transport equations read
\be
\partial_t E_\n + \frac{1}{r^2} \partial_r r^2 F_\n^r &=& \eta_\n -
\kappa_{\n,a} E_\n\,,\\
\partial_t F_\n^r + \frac{1}{r^2} \partial_r r^2 P_\n^{rr} &=& -
(\kappa_{\n,a}+\kappa_{\n,s}) F_\n^r + \, \no \\
 && E_\n\frac{1-p}{r}\,, 
\ee 
\noindent
where $\eta_\n$, $\kappa_{\n,a}$, and $\kappa_{\n,s}$ are the neutrino emissivity, absorption opacity, and scattering opacity, respectively. To numerically solve these equations we discretize the neutrino spectrum into 48 energy groups, logarithmically spaced between 0.5 MeV
and 180 MeV. For each energy group and species, we compute the closure ($p = P_\nu^{rr}/E$), the spatial flux terms ($\partial_r r^2 F_\n^r$ and $\partial_r r^2 P_\n^{rr}$), and the values of the neutrino interaction coefficients ($\eta_\n$, $\kappa_{\n,a}$, and $\kappa_{\n,s}$) explicitly at the beginning of the time step (denoted via the index $(n)$). We then use a first order, implicit/explicit time integration method to solve for the values of the energy and
momentum density at time $t+\Delta t$ (or the $(n+1)$ time step), 
\be
&&E^{(n+1)}_\n = [E^{(n)}_\nu - \Delta t (\partial_r
(r^2F_\n^{r,(n)})/r^2 + \eta_\n)]\times \no \\
&&\hspace*{0.5cm} 1/ (1+\kappa_{\n,a} \Delta
t)\,, \\
\nonumber && F^{r,(n+1)}_\n = \{F^{r,(n)}_\nu - \Delta t [\partial_r
(r^2P_\n^{rr,(n)})/r^2 + \\
&&\hspace*{0.5cm}E^{(n+1)}_\n (1-p_\n^{(n)})/r]\}/
[1+(\kappa_{\n,a}+\kappa_{\n,s}) \Delta t]\,.
\ee
The explicit calculation of the spatial flux remains valid in the diffusion limit due to corrections applied to the Riemann solution in high optical depth regions \citep{oconnor:15a}.  The neutrino interaction coefficients are computed using \code{NuLib}, an open-source neutrino interaction library available at \url{http://www.nulib.org} \citep{oconnor:15a}. For PNS calculations, we include elastic scattering of neutrinos on nucleons, and coherent elastic scattering on alpha particles and heavy nuclei as contributions to $\kappa_{\n,s}$. Charged current absorption of electron neutrinos on neutrons and heavy nuclei and electron antineutrinos on protons is included in $\kappa_{\n,a}$ and in $\eta_\n$ via Kirchhoff's law.  For heavy lepton neutrinos we determine the emissivity $\eta_\n$ from pair processes (electron-positron annihilation and nucleon-nucleon Bremsstrahlung) and an effective absorption opacity via an approximation that works well for supernova conditions \citep{oconnor:15a}.

\subsection{The Monte-Carlo Neutrino Transport Code}

In order to asses the quality of the M1 closures, we compare the M1 results to Monte Carlo radiation transport calculations using the code of \citet{abdikamalov:12}. Here, we outline some salient aspects of such methods, while more in-depth discussion can be found in \citet{abdikamalov:12}.

Monte Carlo methods have been used for many applications \citep[e.g.,][]{janka:89c,janka:91phd,janka:92c,burrows:95,wolf:99,keil:03,lucy:05,kasen:06,densmore:07,dolence:09,wollaeger:13,richers:15}. Such methods use sequences of pseudo-random numbers to simulate the transport of radiation using the concept of Monte Carlo particles. Each MC particle represents a group of physical particles with a given location, direction, and energy (or frequency). The number of physical radiation particles (i.e., photons or neutrinos) represented by a MC particle is called the weight of the MC particle. The smaller the weight, the larger the number of MC particles that are needed to model a given problem, which means higher precision at the price of higher computational cost. 

The spatial problem domain is divided into a number of interconnected cells, and matter in each cell has its own temperature, density, and composition. Using this information, one calculates the number of MC particles that has to be emitted in each cell within a timestep. These particles are then randomly placed in each cell by randomly sampling their coordinates. The propagation direction is sampled randomly with isotropic distribution. The frequencies/energies of particles are chosen using the form of the energy-dependence of the emissivity function. 

Once the MC particles are placed in each cell, they are then transported within a timestep. This can be achieved by calculating three distances for each particle: the distance to collision (absorption or scattering), the distance to the boundary of its cell, and the distance the particle would travel until the end of timestep if it were to travel freely (i.e., assuming that no collision happens). What happens to a particle depends on which of these distances is the smallest. If the distance to collision is the smallest, then the particle is either absorbed or moved to its new location and scattered. The probabilities of absorption and scattering are proportional to their relative opacities. If the distance to a cell boundary is the smallest, then the particle is moved to the boundary of the next cell and transported further in the new cell by calculating a new set of three distances. If the particle crosses the outer boundary of the computational domain, it is removed from the system. Finally, if the third distance is the smallest, then the particle moves by that distance within its cell. Once this step is accomplished for all of the MC particles, the remaining particles, i.e., the ones that have not been absorbed or left the domain, are stored in memory as a preparation for the next timestep. At the next timestep, the cells are populated with newly emitted MC particles, which together with the MC particles remaining from the previous step are then transported further within the second timestep. Subsequent timesteps are performed in the same manner until the end of the simulation.

\bibliography{bibliography/gw_references,bibliography/eos_references,bibliography/methods_references,bibliography/radiation_transport_references,bibliography/bh_formation_references,bibliography/sn_theory_references,bibliography/nu_oscillations_references,bibliography/NSNS_NSBH_references,bibliography/nu_interactions_references,murchikova_references}
\bibliographystyle{mnras.bst}

\bsp
\label{lastpage}
\end{document}